\documentclass[showpacs,aps,pra,twocolumn]{revtex4}
\usepackage{graphicx,epsfig}
\usepackage{times}
\usepackage{amsmath,amssymb,wasysym}
\usepackage{color}

\newlength\figurewidth
\newcommand{\quoting}[1]{``#1''}
\newcommand{\vc}[1]{\vec{#1}}
\newcommand{\rem}[1]{}

\newcommand{\imag}[1]{\text{Im}(#1)}

\newcommand{\imagc}[1]{\text{Im}\,#1}
\newcommand{\real}[1]{\text{Re}(#1)}
\newcommand{\realb}[1]{\text{Re}[#1]}
\newcommand{\realc}[1]{\text{Re}\,#1}
\newcommand{\limacon}{{lima\c con}}

\begin{document}

\title{Nonorthogonal pairs of copropagating optical modes in deformed microdisk cavities}
  \author{Jan Wiersig, Alexander Ebersp\"acher, Jeong-Bo Shim}
  \affiliation{Institut f{\"u}r Theoretische Physik, Universit{\"a}t Magdeburg,
  Postfach 4120, D-39016 Magdeburg, Germany}
  \author{Jung-Wan Ryu}
\affiliation{Department of Physics Education, Pusan National University, Busan 609-735, Korea}
  \author{Susumu Shinohara, Martina Hentschel}
\affiliation{Max-Planck-Institut f\"ur Physik komplexer Systeme, N{\"o}thnitzer
Stra{\ss }e 38, D-01187
Dresden, Germany}
  \author{Henning Schomerus}
  \affiliation{Department of Physics, Lancaster University, Lancaster LA1 4YB, United Kingdom}
\date{\today}
\begin{abstract}
Recently, it has been shown that spiral-shaped microdisk cavities support highly nonorthogonal pairs of copropagating modes with a preferred sense of rotation (spatial chirality) [Wiersig {\it et al.}, Phys. Rev. A {\bf 78}, 053809 (2008)]. Here, we provide numerical evidence which indicates that such pairs are a common feature of deformed microdisk cavities which lack mirror symmetries. In particular, we demonstrate that discontinuities of the cavity boundary such as the notch in the spiral cavity are not needed. We find a quantitative relation between the nonorthogonality and the chirality of the modes which agrees well with the predictions from an effective non-Hermitian Hamiltonian. A comparison to ray-tracing simulations is given.  
\end{abstract}
\maketitle

\section{Introduction}
Optical microcavities allow trapping of photons for a long time $\tau_c$ in very small volumes~\cite{Vahala03}. This enables one to control light and matter at the nano- and microscale, which is important for many applications such as ultralow threshold lasers~\cite{Park04,WGJ09} and single-photon emitters~\cite{Michler2000,PSV02}.
Microdisks~\cite{MLSGPL92,MSJPLHKH07}, microspheres~\cite{CLBRH93,GMM06}, and microtoroids~\cite{IGYM01,AKSV03} support so-called  whispering-gallery modes. Due to total internal reflection of the photons at the boundary of the cavity these modes have very high quality factors $Q = \omega\tau_c$, where $\omega$ is the resonance frequency. 
However, as a consequence of the rotational symmetry of a microdisk the in-plane light emission from these modes is isotropic, which is a disadvantage for many applications. This problem can be solved by deforming the boundary of the cavity~\cite{LSMGPL93,ND97,GCNNSFSC98}. Several shapes for unidirectional light emission have been proposed, for example the spiral cavity~\cite{CTSCKJ03,HK09,HKB09}, the annular cavity~\cite{WH06}, the {\limacon} cavity~\cite{WH08}, the circular disk with a point scatterer~\cite{DMS09}, and the notched ellipse~\cite{WYY10}.

The spiral cavity is a well-studied system~\cite{CTSCKJ03,HK09,HKB09}. In polar coordinates the boundary of this cavity is defined as
$\rho(\phi) = R\left(1-\frac{\varepsilon}{2\pi}\phi\right)$ with deformation parameter $\varepsilon >0 $ and radius $R$ at
$\phi = 0$. The radius jumps back to $R$ at $\phi = 2\pi$, creating a notch.
The spiral cavity appears to be special in the list of studied geometries for two reasons: (i) it lacks any discrete spatial symmetry and (ii) the boundary curve exhibits a singularity. In Refs.~\cite{Wiersig08,WKH08} 
it has been demonstrated that the modes in this open system come in highly nonorthogonal pairs. Moreover, each pair of modes shows a strong spatial chirality, in the sense that both modes have mainly counterclockwise (CCW) propagating components, while the clockwise (CW) component is weak in both modes. (It is important to emphasize that our usage of the term \quoting{chirality} should not be confused with optical activity in chiral media, see, e.g.,~\cite{Lekner96}.)  The appearance of nonorthogonal and chiral modes in the spiral cavity has been traced back to the asymmetric scattering between CW and CCW propagating waves at the notch~\cite{WKH08}.

The aim of the present paper is to show that this effect also appears in cavities without boundary singularities. We consider two different geometries. One is a rather representative example, an asymmetric version of the {\limacon}. The other example is a variant of a curve of constant width~\cite{Gutkin07}. This exotic geometry helps to clarify the role of ray dynamics. Our results indicate that the nonorthogonal and chiral pairs of modes appear in any cavity geometry which is a sufficiently small deformation of the circle and lacks any mirror symmetries. It should be mentioned that in experiments fabrication tolerances introduce small asymmetries quite naturally, see, e.g.,~\cite{SGS10}. In this sense, the asymmetric shapes are generic which is, however, not reflected in the list of geometries studied in the literature.

This paper is organized as follows. Section~\ref{sec:limacon} reports our numerical results on the properties of optical modes in the asymmetric {\limacon} cavity. In Sec.~\ref{sec:effH} we introduce an effective non-Hermitian Hamiltonian which describes the relation between nonorthogonality and chirality. The ray dynamics in the asymmetric {\limacon} cavity is presented in Sec.~\ref{sec:ray}. In Sec.~\ref{sec:constantwidth} we discuss the cavity with a boundary curve of constant width. We summarize our results in Sec.~\ref{sec:summary}.

\section{Modes in the asymmetric {\limacon}}
\label{sec:limacon}
In the case of (deformed) microdisk cavities with a piece-wise constant effective index of refraction $n(x,y)$, Maxwell's equations can be reduced to a two-dimensional scalar mode equation~\cite{Jackson83eng}
\begin{equation}\label{eq:wave} -\nabla^2\psi =
n^2(x,y)\frac{\omega^2}{c^2}\psi \ ,
\end{equation}
where $\omega = ck$ is the frequency, $k$ is the wave number in vacuum (outside the microdisk), and $c$ is the speed of light in vacuum. The mode equation~(\ref{eq:wave}) is valid for both transverse magnetic (TM) and transverse electric (TE) polarization. For TM polarization the electric field $\vec{E}(x,y,t) \propto (0,0,\realb{\psi(x,y)e^{-i\omega t}})$ is perpendicular to the cavity plane.
The wave function~$\psi$ and its normal derivative~$\partial_\nu\psi$
are continuous across the boundary of the cavity. For TE
polarization, $\psi$ represents the $z$-component of the magnetic field vector
$H_z$. Here, the wave function $\psi$ and $n(x,y)^{-2}\partial_\nu\psi$ are continuous across the boundaries~\cite{Jackson83eng}.
At infinity, outgoing wave conditions are imposed, which results in quasibound
states with complex
frequencies $\omega$ in the lower half-plane. The real part is the
usual frequency and the imaginary part is related to the lifetime
$\tau_c=-1/[2\,\imagc{\omega}]$ and to the quality factor
$Q = -\realc{\omega}/[2\,\imagc{\omega}]$.

\begin{figure}[ht]
\includegraphics[width=\figurewidth]{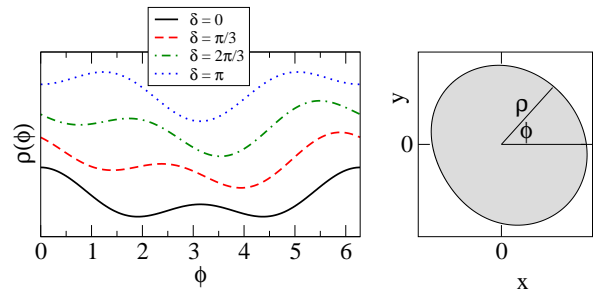}
\caption{(Color online) The left panel shows the boundary parametrization $\rho(\phi)$ in Eq.~(\ref{eq:boundary}) with $\varepsilon_1 = 0.1$ and $\varepsilon_2 = 0.075$, for different values of $\delta$. The curves are shifted vertically for better comparison. The right panel shows a top view of the cavity with $\delta = \pi\frac{\sqrt{5}-1}{2}$. }
\label{fig:sketch}
\end{figure}

In polar coordinates the boundary shape studied in this section is given by 
\begin{equation}\label{eq:boundary}
\rho(\phi) = R\left[1+\varepsilon_1\cos\phi+\varepsilon_2\cos(2\phi+\delta)\right] 
\end{equation}
with $\delta\in[0,2\pi)$. This is illustrated in Fig.~\ref{fig:sketch}. The special case $\varepsilon_2 = 0$ is the {\limacon} cavity~\cite{WH08,SCL08,SHH09,YWD09,WYD09,YKK09,WUS10,SGS10}. For nonzero $\varepsilon_1$, $\varepsilon_2$ and $\delta \neq 0, \pi$ the system does not possess any mirror symmetry. We therefore call the cavity described by Eq.~(\ref{eq:boundary}) the {\it asymmetric {\limacon}}. The ray dynamics in such a geometry is mainly chaotic as in the case of the symmetric {\limacon}. The asymmetric version can be considered as a typical smooth deformation of a circular disk, as the sum of the first three terms of a Fourier expansion of an arbitrary periodic function can be written as in Eq.~(\ref{eq:boundary}).

If not stated otherwise, in the following we use $\delta = \pi\frac{\sqrt{5}-1}{2} \approx 0.618\pi$, $\varepsilon_1 = 0.1$, and $\varepsilon_2 = 0.075$. We have chosen the golden ratio for $\delta/\pi$ to ensure that we are not too close to the symmetric situations $\delta = 0$ and $\pi$. The effective index of refraction is set to $n=3.3$ (e.g. GaAs) and we only consider TM polarization. To compute the complex frequencies and wave functions of quasibound states we use the boundary element method~\cite{Wiersig02b} and, for comparison, the wave matching method, see, e.g.,~\cite{HR02}. We always find very good agreement between these two methods.

\begin{figure}[ht]
\includegraphics[width=\figurewidth]{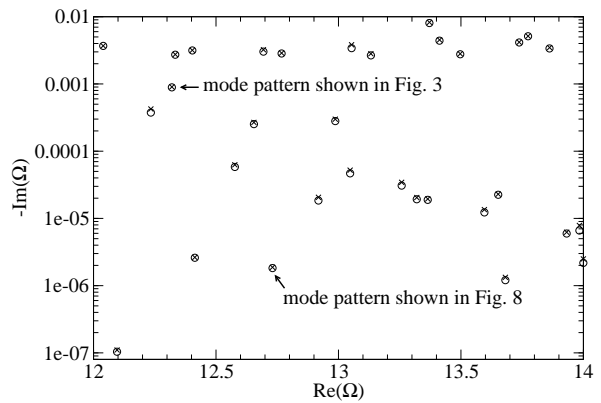}
\caption{Position of dimensionless complex resonance frequencies
with TM polarization for the cavity in Eq.~(\ref{eq:boundary}) with 
refractive index $n=3.3$ and shape parameters $\varepsilon_1 = 0.1$, $\varepsilon_2 = 0.075$, and $\delta = \pi\frac{\sqrt{5}-1}{2}$. Open circles (crosses) mark the slightly higher-$Q$ (lower-$Q$) mode of a given pair of modes.}
\label{fig:frequency_plane}
\end{figure}

Figure~\ref{fig:frequency_plane} shows the resonances in a typical region of the complex plane of normalized frequencies $\Omega = \omega R/c = kR$. It can be clearly seen that modes here always appear in nearly degenerate pairs, even though the deformation of the cavity boundary is not that small. This is due to the rather weak coupling of CW and CCW propagating waves in these open disk-like cavities. One could guess that one member of such a pair is a CW-propagating mode and the other one is a CCW-propagating mode. But this is not true, as we will see in the following. 

\begin{figure}[ht]
\includegraphics[width=\figurewidth]{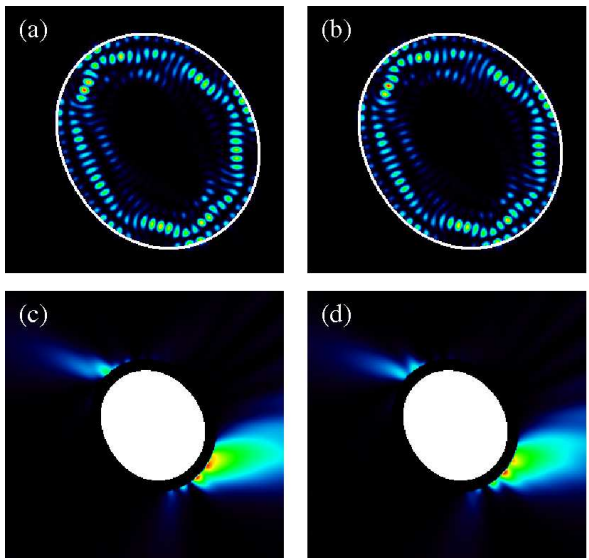}
\caption{(Color online) Intensity $|\psi|^2$ of the nearly degenerate pair of modes in the asymmetric {\limacon} with (a) $\Omega_1 = 12.319807-i0.00089$ and (b) $\Omega_2 = 12.319851-i0.0009$; cf. Fig.~\ref{fig:frequency_plane}. The same color map has been used in panels (a) and (b). Panels (c) and (d) shows the corresponding exterior mode pattern at some small distance away from the cavity (white region). The far-field patterns are shown in Fig.~\ref{fig:farfield2}.}
\label{fig:mode2}
\end{figure}

\begin{figure}[ht]
\includegraphics[width=\figurewidth]{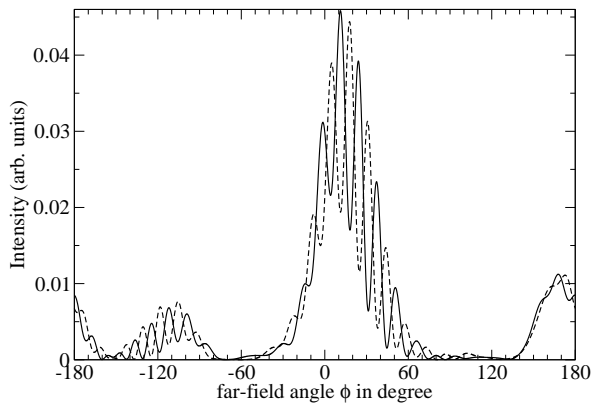}
\caption{Far-field patterns of the modes in
  Fig.~\ref{fig:mode2}. The solid and dashed lines correspond to the modes in  Figs.~\ref{fig:mode2}(a) and~\ref{fig:mode2}(b), respectively. For the definition of the far-field angle $\phi$ see Fig.~\ref{fig:sketch}.} \label{fig:farfield2}
\end{figure}

Figure~\ref{fig:mode2} depicts a typical example of such a pair of nearly degenerate modes. The splitting in real part $\Delta \real{\Omega} \approx 4.4\times 10^{-5}$ and imaginary part  $\Delta \imag{\Omega} \approx 10^{-5}$ is very small. The spatial mode pattern is difficult to distinguish by eye. A closer look at the far-field pattern shown in Fig.~\ref{fig:farfield2} reveals that they have the same envelope, but there clearly are different oscillations on top of this envelope. Note that not only the far-field pattern indicates unidirectional light emission, the near-field pattern in Figs.~\ref{fig:mode2}(c) and~\ref{fig:mode2}(d) shows that the emission follows a single-lobe beam, which is interesting for applications.

Following Refs.~\cite{Wiersig08,WKH08} we analyze the mode pattern by
expanding the wave function inside the cavity in cylindrical harmonics,
\begin{equation}\label{eq:amd}
\psi(\rho,\phi) = \sum_{m=-\infty}^{\infty} \alpha_m J_m(nk\rho)\exp{(im\phi)} \ ,
\end{equation}
where $J_m$ is the $m$th order Bessel function of the first kind. Positive 
(negative) values of the angular momentum index $m$ correspond to CCW (CW) 
traveling-wave components. As the origin of this expansion we choose the center of mass of the cavity, $(x,y) = (\varepsilon_1 R/2,0)$. Note that this particular choice does not affect our conclusions (as long as the origin is chosen inside the cavity). 
The coefficients in the expansion in Eq.~(\ref{eq:amd}) are naturally given in the case of the wave matching method~\cite{HR02}. In the case of the boundary element method~\cite{Wiersig02b} we use a Fourier transformation of the wave function to determine the coefficients. Both approaches give identical numerical results.

\begin{figure}[ht]
\includegraphics[width=\figurewidth]{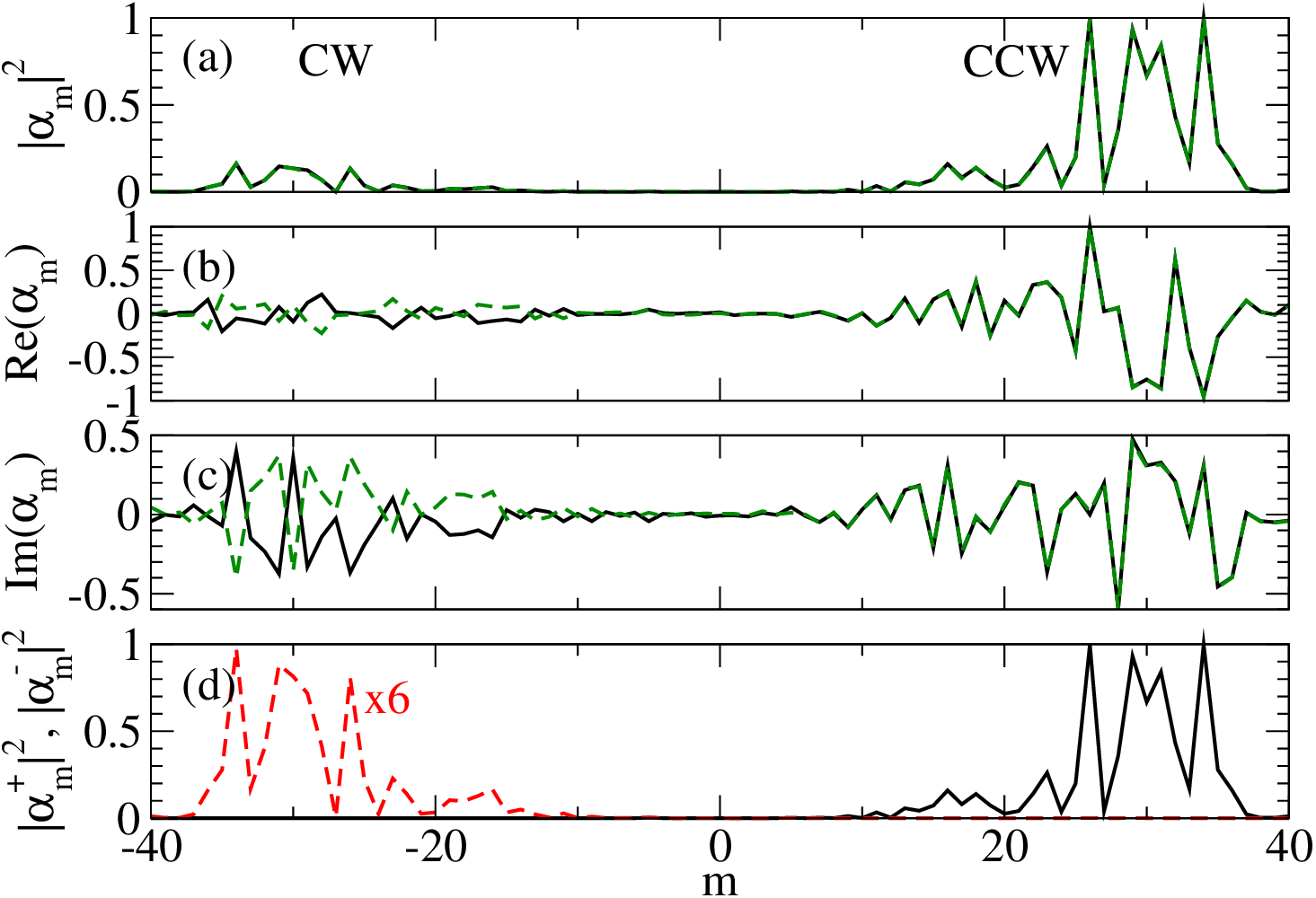}
\caption{(Color online) Angular momentum distributions $\alpha^{(1)}_m$ (black solid line) and $\alpha^{(2)}_m$ (green dashed) of the modes in Fig.~\ref{fig:mode2} (real part normalized to 1 at maximum): (a) absolute value squared, (b) real and (c) imaginary part, (d) superpositions $\alpha^+_m =
 (\alpha^{(1)}_m+ \alpha^{(2)}_m)/2$ (black solid) and $\alpha^-_m
= (\alpha^{(1)}_m-\alpha^{(2)}_m)/2$ (red dashed, multiplied by a factor of~6).}
\label{fig:AMD2}
\end{figure}

In Fig.~\ref{fig:AMD2}(a) we can observe that for both modes the angular momentum
distribution $|\alpha_m|^2$ is dominated by the CCW component, i.e., none of
the two modes can be classified as a CW traveling-wave mode. The small difference
between the expansion coefficients of the modes can be seen in Figs.~\ref{fig:AMD2}(b) and~\ref{fig:AMD2}(c). For negative
angular momentum index both the real and the imaginary part of $\alpha_m$ have
a different sign for the two modes. That means that we can
construct superpositions with $\alpha^\pm_m = (\alpha^{(1)}_m\pm
\alpha^{(2)}_m)/2$ being CW and CCW traveling-waves, respectively, as can be
seen in Fig.~\ref{fig:AMD2}(d). However, the CCW superposition has a much larger amplitude. It is important to emphasize that these superpositions are not eigenmodes of the cavity as they are composed of two modes with slightly different frequencies and $Q$-factors.

It is convenient to use the angular momentum representation~(\ref{eq:amd}) to define the (spatial) chirality of a mode by
\begin{equation}\label{eq:chirality}
\alpha = 1-\frac{\min\left(\sum_{m=-\infty}^{-1}|\alpha_m|^2,\sum_{m=1}^{\infty}|\alpha_m|^2\right)}{\max\left(\sum_{m=-\infty}^{-1}|\alpha_m|^2,\sum_{m=1}^{\infty}|\alpha_m|^2\right)} \ .
\end{equation}
If the weight of the CW and CCW components is equally distributed then the chirality is $\alpha=0$. This is for instance the case for a cavity which possesses a mirror symmetry. To see this, choose the coordinate system such that $\rho(-\phi) = \rho(\phi)$. In the angular momentum representation~(\ref{eq:amd}) modes with positive (negative) parity $\psi(\rho,-\phi) = \pm\psi(\rho,\phi)$ must have $a_{-m}(-1)^m=\pm a_m$. In both cases, $|a_{-m}|^2=|a_m|^2$ which according to Eq.~(\ref{eq:chirality}) gives a chirality~$\alpha = 0$. The same is true in the case of a closed system (in nonlinear dynamics a closed cavity  is called a billiard) with real-valued frequencies, where $a^*_{-m}(-1)^m = a_m$ which again leads to $|a_{-m}|^2=|a_m|^2$. Note that the former statement is correct for the special case of the circular microcavity with its degenerate pairs of modes only if the linear superpositions leading to standing waves are chosen.
The other extreme of full chirality, $\alpha = 1$, is realized if a mode has no CW (or CCW) component at all. 

The asymmetric {\limacon} (with $\delta \neq 0, \pi$) lacks any mirror symmetry. Moreover, the system is open and has complex-valued frequencies due to the outgoing-wave conditions at infinity. For the modes in Figs.~\ref{fig:mode2}(a) and~\ref{fig:mode2}(b) we find numerically $\alpha \approx 0.839$ and $\alpha \approx 0.8404$, respectively. Hence, both modes show a strong chirality.

Now we demonstrate that the modes not only have a strong chirality but that they are also pairwise highly nonorthogonal. To quantify the nonorthogonality we compute the normalized overlap integral of two modes $\psi_1$ and $\psi_2$ over the interior of the cavity~${\cal C}$ (see also, e.g.,~\cite{CM98})
\begin{equation}
\label{eq:overlap}
S = \frac{|\int_{\cal C} dxdy\;\psi_1^*\psi_2|}{\sqrt{\int_{\cal C} dxdy\;\psi_1^*\psi_1}\sqrt{\int_{\cal C} dxdy\;\psi_2^*\psi_2}} \ .
\end{equation}
In the case of orthogonal states $S=0$ and in the case of collinear states $S=1$. It is easy to show that in the presence of a mirror symmetry or for a closed system the overlap~$S$ vanishes.
For the pair of modes in Fig.~\ref{fig:mode2}, however, we find $S\approx 0.7236$ reflecting a strong nonorthogonality. Mode nonorthogonality is important as it implies excess quantum noise~\cite{Siegman89a,Siegman89b,SFP00,LRS08,Schomerus09}.

\begin{figure}
\includegraphics[width=\figurewidth]{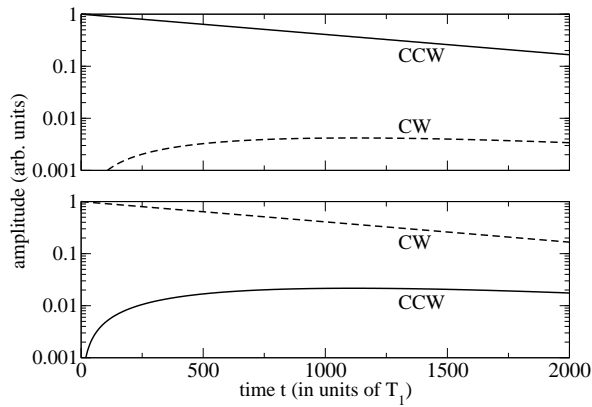}
\caption{Time evolution of CCW (solid lines) and CW (dashed) components in semilogarithmic scale. The traveling waves are superpositions of the modes with frequencies $\Omega_1$ and $\Omega_2$ depicted in Fig.~\ref{fig:mode2}. The upper (lower) panel contains the dynamics starting with a pure CCW (CW) traveling wave. Time is measured in units of $T_1 = 2\pi/\realc{\Omega_1}$.} \label{fig:dynamicsCCW}
\end{figure}

One might think that the strong chirality must have a large influence on the dynamics of waves in such a cavity. For the asymmetric {\limacon} the impact is, however, weak. The upper panel of Fig.~\ref{fig:dynamicsCCW} shows the dynamics of the CCW and CW components of an initially (at time $t=0$) CCW traveling wave using proper superpositions of the modes shown in Fig.~\ref{fig:mode2} (we follow the procedure explained in detail in Ref.~\cite{WKH08}). Only a very weak scattering into the CW component can be observed. The lower panel of Fig.~\ref{fig:dynamicsCCW} shows the situation where the initial wave is purely CW propagating. Here, the backscattering into the CCW component is significantly larger, though still weak. 

\begin{figure}[ht]
\includegraphics[width=\figurewidth]{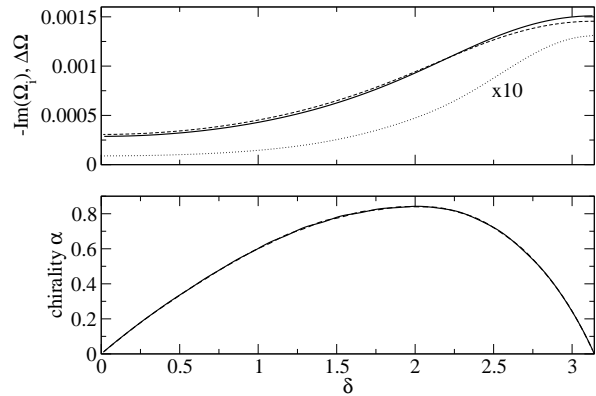}
\caption{Upper panel: individual decay rates $-\imag{\Omega_i}$ (solid and dashed line) and the level splitting $\Delta\Omega = |\real{\Omega_1}-\real{\Omega_2}|$ (dotted line, scaled by a factor of 10) of the pair of modes in Fig.~\ref{fig:mode2} vs shape parameter~$\delta$. Lower panel: corresponding chirality $\alpha$ as function of~$\delta$. Note that two curves are on top of each other.}
\label{fig:dynamics}
\end{figure}

The chirality of the modes in Fig.~\ref{fig:mode2} as a function of the asymmetry parameter~$\delta$ in the interval $[0,\pi]$ is shown in the lower panel of Fig.~\ref{fig:dynamics}. 
The curves have been computed by starting with $\delta = \pi\frac{\sqrt{5}-1}{2}$, decreasing $\delta$ in small steps, and thereby following the modes. Second, $\delta$ has been varied between $\pi\frac{\sqrt{5}-1}{2}$ and $\pi$. 200 discretization points on the $\delta$-axis have been used for each mode. As the total change in the frequency is rather small in this range of parameter variation we do not observe any avoided resonance crossing.
For $\delta = 0$ and $\delta = \pi$ the system possess a mirror symmetry and therefore we observe no chirality, $\alpha = 0$. The maximum chirality of about $0.845$ is attained at $\delta \approx 2$, which is close to the value that we mainly use in this paper, $\delta = \pi\frac{\sqrt{5}-1}{2} \approx 1.94$. The upper panel shows the level splitting $\Delta\Omega = |\real{\Omega_1}-\real{\Omega_2}|$ and the individual decay rates of the two modes $-\imag{\Omega_i}$ as a function of the asymmetry parameter~$\delta$. Note that the level splitting is always much smaller than the individual decay rates, i.e., the spectral width of the two modes strongly overlap.

\begin{figure}[ht]
\includegraphics[width=\figurewidth]{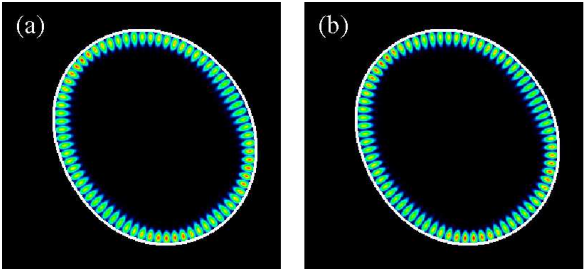}
\caption{(Color online) Intensity $|\psi|^2$ of the nearly degenerate pair of modes in the asymmetric {\limacon} with (a) $\Omega_1 = 12.73070292-i1.83\times 10^{-6}$ and (b) $\Omega_2 = 12.73070286-i1.88\times 10^{-6}$; cf. Fig.~\ref{fig:frequency_plane}.}
\label{fig:mode1}
\end{figure}

\begin{figure}[ht]
\includegraphics[width=\figurewidth]{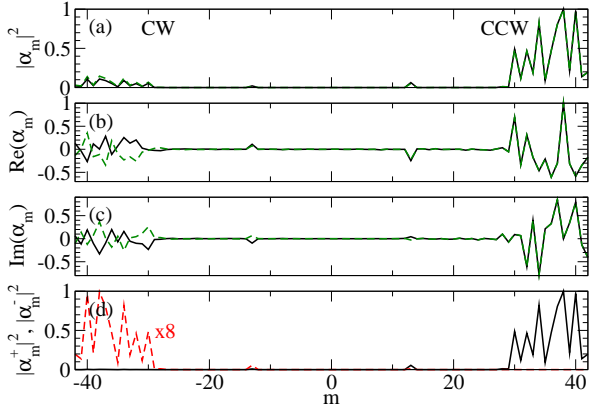}
\caption{(Color online) Angular momentum distributions $\alpha^{(1)}_m$ (black solid line) and $\alpha^{(2)}_m$ (green dashed) of the modes in Fig.~\ref{fig:mode1} normalized to 1 at maximum: (a) absolute
value squared, (b) real and (c) imaginary parts, and (d) superpositions $\alpha^+_m =
 (\alpha^{(1)}_m+ \alpha^{(2)}_m)/2$ (black solid) and $\alpha^-_m
= (\alpha^{(1)}_m-\alpha^{(2)}_m)/2$ (red dashed, multiplied by a factor of 8).}
\label{fig:AMD1}
\end{figure}

Another example of a pair of nearly degenerate modes is shown in Figs.~\ref{fig:mode1} and \ref{fig:AMD1}. Again we find that both modes exhibit a strong chirality ($\alpha \approx 0.8793$ and $\alpha \approx 0.8658$) and a significant nonorthogonality ($S\approx 0.7778$). This is in particular remarkable as the quality factor of the modes is about $3.5\times 10^{6}$. For such enormously high quality factors one would expect a behavior similar to that of orthogonal states in a closed system such as in a billiard. This reasoning is, however, too naive since the frequency splitting is much less than the individual decay rates. The resonances therefore strongly overlap, which is usually considered as a feature associated with strongly open systems. 

It is also important to mention that the overlap of long-lived modes from different pairs is significantly smaller. For instance, for one mode in Fig.~\ref{fig:mode2} and one from  Fig.~\ref{fig:mode1} we always find $S < 6\times 10^{-4}$. That means the nonorthogonality is significant only within each pair of long-lived modes. 
\rem{
\begin{figure}[ht]
\includegraphics[width=\figurewidth]{ff1.eps}
\caption{Far-field patterns of the modes in
  Fig.~\ref{fig:mode1}. The solid and dashed lines correspond to Fig.~\ref{fig:mode1}(a) and (b), respectively. For the definition of the far-field angle $\phi$ see Fig.~\ref{fig:sketch}.} \label{fig:farfield1}
\end{figure}
}

\begin{figure}[ht]
\includegraphics[width=\figurewidth]{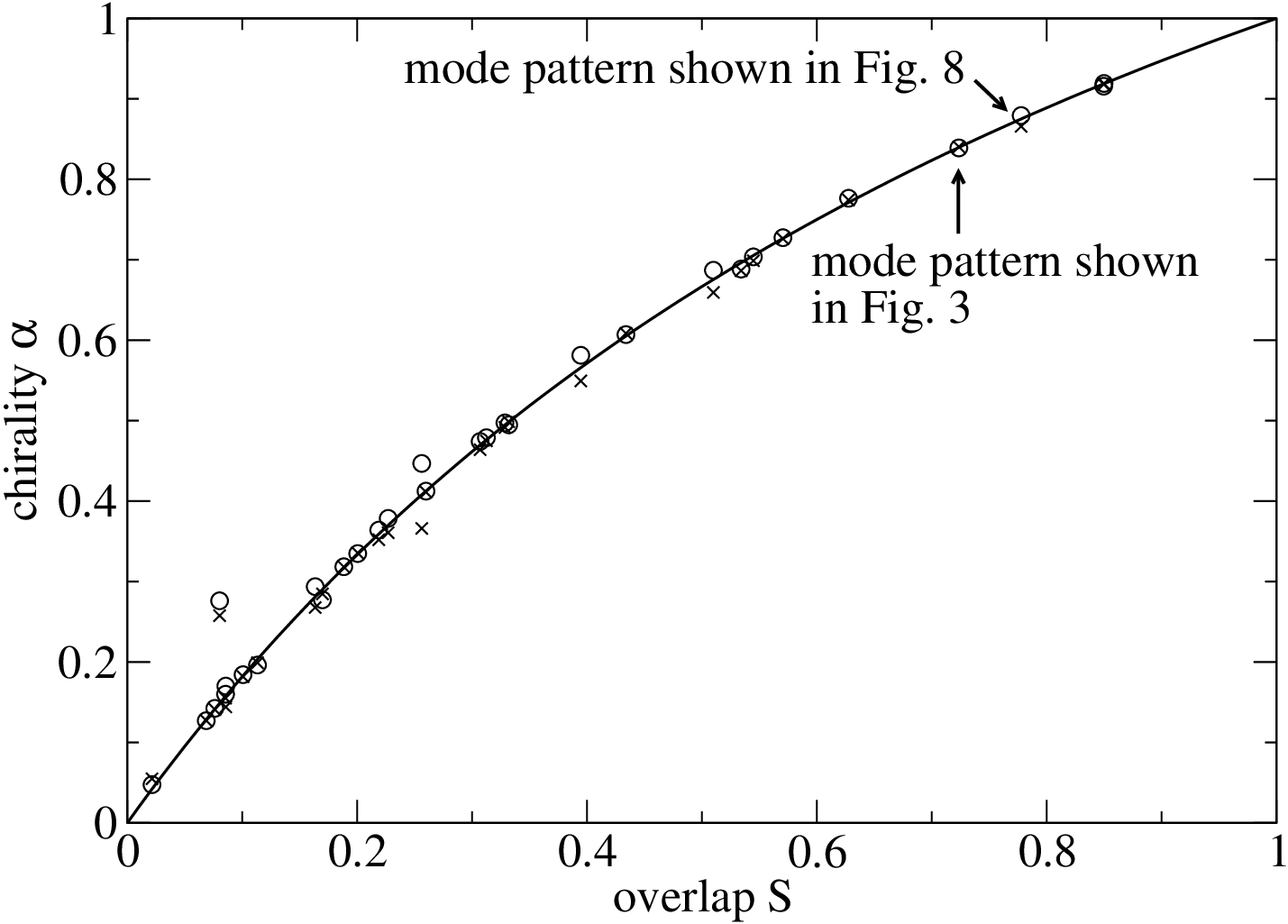}
\caption{Chirality $\alpha$ vs spatial overlap $S$ of pairs of almost degenerate modes in the asymmetric {\limacon}. Open circles (crosses) mark the slightly higher-$Q$ (lower-$Q$) mode of a given pair of modes computed numerically from Maxwell's equations; cf. Fig.~\ref{fig:frequency_plane}. The solid line is the analytical prediction of the theoretical model, Eq.~(\ref{eq:chirality2by2}).}
\label{fig:chirality}
\end{figure}

Figure~\ref{fig:chirality} summarizes the results on the chirality and the pairwise nonorthogonality of the modes which are present in the frequency regime considered in Fig.~\ref{fig:frequency_plane}. The mode pair with the highest $Q$-factors ($\Omega_1 \approx 12.0960925-i 1.04\times 10^{-7}$ and $\Omega_2 \approx 12.09609251-i 1.11\times 10^{-7}$) is, however, not shown here. In this (and only in this) case the numerical computation of the chirality is not fully converged due to the exceptionally strong degeneracy of this particular mode pair even though we use up to $32\,000$ discretization points in the boundary element method. From Fig.~\ref{fig:chirality} it can be observed that the chirality and the overlap are correlated. This correlation can be explained by an effective  non-Hermitian Hamiltonian, which is discussed in the next section. 
Finally, we remark that 29 of the 31 considered pairs of modes exhibit a larger CCW component. Only two pairs have a larger CW component. 

For smaller frequencies the chirality and the nonorthogonality are weaker (not shown). An intuitive explanation is that for very small frequencies, i.e., large wavelengths, the modes do not feel the asymmetry of the boundary shape anymore.

\section{Effective non-Hermitian Hamiltonian}
\label{sec:effH}
Reference~\cite{WKH08} introduced a simple toy model to describe the main features of the chirality and nonorthogonality of modes in the spiral cavity. Here we use the same two-by-two non-Hermitian and nonsymmetric matrix
\begin{equation}\label{eq:nonHermitianMatrix}
H = \left(\begin{array}{cc} \omega_0 & 0 \\ 0 & \omega_0 \end{array}\right) +
    \left(\begin{array}{cc} \Gamma & V \\ \eta V^* & \Gamma \end{array}\right) \ .
\end{equation}
For the convenience of the discussion, our interpretation is that this matrix describes the dynamics of the wave function~$\psi$ in slowly-varying envelope approximation~\cite{Siegman86} in the time domain by a Schr\"odinger-type equation
\begin{equation}\label{eq:slowlyvarying}
i\frac{\partial}{\partial t}\psi = H\psi \ .
\end{equation}
When deriving this equation from the Maxwell's equations one assumes that the optical field varies slowly in time (not necessarily in space) with respect to a reference frequency which we choose to be close to the two nearly degenerate modes of interest. 

The nonsymmetric Hamiltonian matrix~(\ref{eq:nonHermitianMatrix}) is defined in the CCW/CW traveling-wave basis
\begin{equation}\label{eq:travelingwave}
\vc{t}_1 = \left(\begin{array}{c} 1 \\ 0\end{array}\right)
\,;\quad
\vc{t}_2 = \left(\begin{array}{c} 0 \\ 1\end{array}\right) \ .
\end{equation}
The eigenvectors of the first matrix in Eq.~(\ref{eq:nonHermitianMatrix}) on the right-hand side belong to the CCW and CW
traveling waves with, for simplicity, equal frequency $\omega_0\in\mathbb{C}$ in the
absence of any coupling between them. The second matrix accounts for coupling of CCW and CW traveling components. The diagonal elements are given by the total decay rates and frequency shifts $\Gamma\in\mathbb{C}$ which are assumed to be equal for simplicity. The off-diagonal element $V = |V|e^{i\beta} \in\mathbb{C}$ describes scattering from a CW traveling wave into the CCW traveling wave. The other off-diagonal element $\eta V^*$ describes scattering from a CCW traveling wave into the CW traveling wave. The latter scattering is assumed to be weaker, i.e., $|\eta|<1$. Therefore, here $|\eta|$ plays the role of the asymmetry parameter.

Note that a standing wave basis can be chosen as 
\begin{equation}\label{eq:standingwave}
\vc{s}_1 = \frac{1}{\sqrt{2}}\left(\begin{array}{c} 1 \\ 1\end{array}\right)
\,;\quad
\vc{s}_2 =  \frac{i}{\sqrt{2}}\left(\begin{array}{c} -1 \\ 1\end{array}\right) \ .
\end{equation}
While the traveling-wave basis~(\ref{eq:travelingwave}) corresponds to terms $e^{im\phi}$ with $m>0$ and $m < 0$ in the angular momentum representation~(\ref{eq:amd}), the standing wave basis corresponds to $\cos{(m\phi)}$ and $\sin{(m\phi)}$ with nonnegative $m$. In the standing wave basis the non-Hermitian Hamiltonian matrix is symmetric (as required by time-reversal symmetry):
\begin{equation}
H = \left(\begin{array}{cc} \omega_0+\Gamma+\frac{V+\eta V^*}{2} & \frac{i}{2}(V-\eta V^*)\\ \frac{i}{2}(V-\eta V^*) & \omega_0+\Gamma-\frac{V+\eta V^*}{2}\end{array}\right) \ .
\end{equation}
    
The complex eigenvalues of the matrix~(\ref{eq:nonHermitianMatrix}) are given by
\begin{equation}\label{eq:eigenvalue}
\omega_\pm = \omega_0+\Gamma\pm\sqrt{\eta}|V| \ . 
\end{equation}
The (not normalized) right-hand eigenvectors in the CCW/CW traveling-wave basis turn out to be
\begin{equation}
\vec{\alpha}_\pm = \left(\begin{array}{c} 1 \\ \pm\sqrt{\eta}e^{-i\beta}\end{array}\right) \ . \label{eq:eigenvector}
\end{equation}
These eigenvectors explain the mode structure observed in
Sec.~\ref{sec:limacon} including the sign difference in the CW components as well as the relative weight of the CCW and CW components. The weight of the first component (corresponding to CCW traveling waves) squared $\sim 1$ is much larger than that of the second 
component (corresponding to CW traveling waves) squared $\sim |\eta|$, cf. Figs.~\ref{fig:AMD2} and~\ref{fig:AMD1}. Hence, the $2\times 2$ model predicts an identical chirality for both modes,
\begin{equation}\label{eq:chiralityalpha}
\alpha = 1-|\eta| \ ,
\end{equation}
which is nonzero in the case of asymmetric coupling ($|\eta|\neq 1$). Note that the chirality~$\alpha$ does not depend on the coupling strength $|V|$. As a consequence, even an infinitesimal coupling can lead to a significant chirality. The reason behind this singular behavior is the two-fold degeneracy of the unperturbed modes.

The two eigenvectors in Eq.~(\ref{eq:eigenvector}) are, in general, nonorthogonal, i.e., the normalized overlap             
\begin{equation}\label{eq:nonortho}
S = \frac{|\vec{\alpha}_+^*\cdot\vec{\alpha}_-|}{|\vec{\alpha}_+||\vec{\alpha}_-|} = \frac{1-|\eta|}{1+|\eta|} 
\end{equation}
does not vanish. Using this result and Eq.~(\ref{eq:chiralityalpha}) we arrive at a relation between the chirality of the two modes and their overlap,
\begin{equation}\label{eq:chirality2by2}
\alpha = \frac{2S}{1+S} \ .
\end{equation}
Figure~\ref{fig:chirality} compares this prediction with the data obtained from numerical solutions of Maxwell's equations. It can be seen that the $2\times 2$ model works very well. 
\rem{
Only the two pairs of modes at $S\approx 0.259$ and $S\approx 0.362$ show a significant deviation from the relation in Eq.~(\ref{eq:chirality2by2}). In these two particular cases the numerical methods to compute the modes are not fully converged because of an exceptionally strong degeneracy of the modes combined with a high $Q$-factor (very small value of $\imag{kR}$). We, therefore, believe that the observed deviations from the prediction of the $2\times 2$ model is an numerical artefact.
}

\begin{figure}[ht]
\includegraphics[width=\figurewidth]{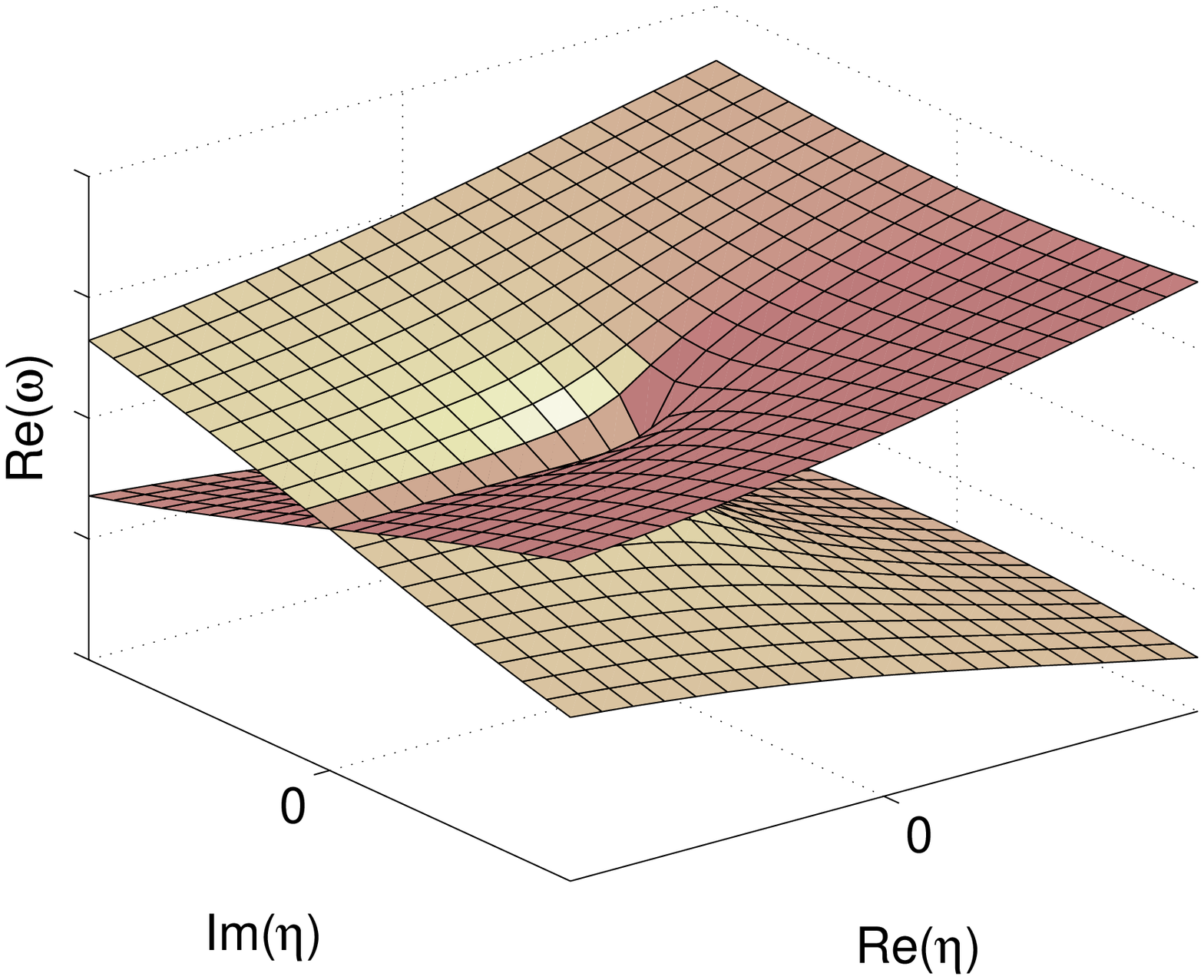}
\caption{(Color online) Complex-square-root topology with a branch point singularity at the exceptional point of the Hamiltonian~(\ref{eq:nonHermitianMatrix}).}
\label{fig:ReE}
\end{figure}

For the case $\eta = 0$, the Hamiltonian~(\ref{eq:nonHermitianMatrix}) exhibits an exceptional point~\cite{Heiss00,LRS08}, i.e., not only the eigenvalues become degenerate,
\begin{equation}
\omega_\pm = \omega_0+\Gamma \ ,
\end{equation}
but also the eigenvectors collapse to a single one, which in the CCW/CW traveling-wave basis reads
\begin{equation}
\vec{\alpha}_\pm = \left(\begin{array}{c} 1 \\ 0\end{array}\right) \ . 
\end{equation}
The corresponding complex-square-root topology of the eigenvalues~(\ref{eq:eigenvalue}) with a branch point singularity is shown in Fig.~\ref{fig:ReE}.
The eigenvector at the exceptional point in the standing wave basis~(\ref{eq:standingwave})  is given by
\begin{equation}
\vec{\alpha}_\pm \sim  \left(\begin{array}{c} 1 \\ 0\end{array}\right) +i \left(\begin{array}{c} 0 \\ 1\end{array}\right) \ .
\end{equation}
This is a chiral state in the sense of Refs.~\cite{DDG03,HH01}.

The solutions $(\psi_1(t),\psi_2(t))$ of the Schr\"odinger-type equation~(\ref{eq:slowlyvarying}) can be found analytically. We consider the time scale related to the strength of the mode splitting
\begin{equation}\label{eq:tauex}
T = \frac{1}{\sqrt{|\eta|}|V|}
\end{equation}
and the decay time
\begin{equation}\label{eq:taudecay}
\tau = -\frac{1}{\imag{\omega_0+\Gamma}} \ .
\end{equation}
In the following we restrict ourselves to the case $\tau\ll T$ and $t\ll T$ which is the relevant regime for small $|\eta|$. 
We find for a wave propagating initially at $t=0$ in CCW direction with normalized amplitude
\begin{eqnarray}\label{eq:psi12a}
|\psi_1| & = & e^{-t/\tau}\ ,\\
\label{eq:psi12b}
|\psi_2| & = & \sqrt{|\eta|}\frac{t}{T}e^{-t/\tau}\ .
\end{eqnarray}
The component $|\psi_1|$ decays from 1 to 0 in an exponential manner and $|\psi_2|$ increases from zero to the value
\begin{equation}\label{eq:max_small}
|\psi_2|_{\mbox{\footnotesize max}} = \frac{\sqrt{|\eta|}\tau}{T}e^{-1} 
\end{equation}
and then decays to zero. For $|\eta| \ll 1$ the value of $|\psi_2|_{\mbox{\footnotesize max}}$ is very small. This corresponds to the situation in the upper panel of Fig.~\ref{fig:dynamicsCCW}. Note that Eqs.~(\ref{eq:psi12a})-(\ref{eq:psi12b}) are exact for $\eta=0$, i.e., at the exceptional point; see also~\cite{WKH08,Heiss10}.

For a wave propagating initially at $t=0$ in CW direction it follows that
\begin{eqnarray}
|\psi_1| & = & \frac{1}{\sqrt{|\eta|}}\frac{t}{T}e^{-t/\tau}\ ,\\
|\psi_2| & = & e^{-t/\tau}\ .
\end{eqnarray}
The component $|\psi_2|$ decays from 1 to 0 in an exponential manner. However, $|\psi_1|$ increases from zero to the value
\begin{equation}\label{eq:max}
|\psi_1|_{\mbox{\footnotesize max}} = \frac{\tau}{\sqrt{|\eta|}T}e^{-1} 
\end{equation}
and then decays to zero. Here, $|\psi_1|_{\mbox{\footnotesize max}}$, in principle, can be close to 1, as $|\eta|$ is a small number. The ratio
\begin{equation}
\frac{|\psi_1|_{\mbox{\footnotesize max}}}{|\psi_2|_{\mbox{\footnotesize max}}} = \frac{1}{|\eta|} 
\end{equation}
reflects the fact that the scattering from CW to CCW is $1/|\eta|$ times stronger than the scattering from CCW to CW. 

For the modes in the asymmetric {\limacon} in Fig.~\ref{fig:mode2} we estimate $\tau/T \approx 0.024$, from Fig.~\ref{fig:AMD2} $|\eta| \approx 1/6$, and therefore 
$|\psi_1|_{\mbox{\footnotesize max}} \approx 0.022$ and $|\psi_2|_{\mbox{\footnotesize max}} \approx 0.0037$, in reasonable agreement with Fig.~\ref{fig:dynamicsCCW}. For the modes in Fig.~\ref{fig:mode1} we estimate $|\psi_1|_{\mbox{\footnotesize max}} \approx 0.017$ and $|\psi_2|_{\mbox{\footnotesize max}} \approx 0.0021$. If we apply the same analysis to the spiral cavity then we find that the $|\psi_1|_{\mbox{\footnotesize max}}$-values are significantly larger. This is consistent with the fact that the mode splitting and the chirality in the spiral cavity is larger.

\section{Ray dynamics in the asymmetric {\limacon}}
\label{sec:ray}
\begin{figure}[ht]
\includegraphics[width=\figurewidth]{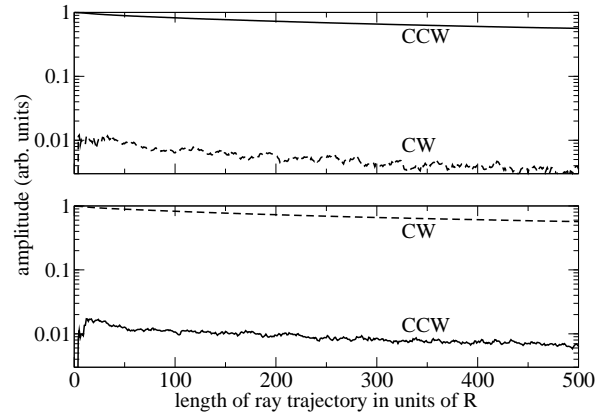}
\caption{Time evolution of amplitude (defined as square root of intensity) in semilogarithmic scale corresponding to CCW (solid lines) and CW (dashed) propagating light rays in the asymmetric {\limacon}. The upper (lower) panel shows the dynamics starting with a set of pure CCW (CW) propagating rays in analogy to the wave dynamical considerations in Fig.~\ref{fig:dynamicsCCW}. Time is proportional to the geometric length of ray trajectories.} \label{fig:rays}
\end{figure}

Figure~\ref{fig:rays} shows the ray analog of Fig.~\ref{fig:dynamicsCCW}. In the upper (lower) panel a bunch of  $12\,500$ rays in the asymmetric {\limacon} propagating in CCW (CW) direction with angle of incidence $|\chi|$ distributed uniformly well above the critical angle for total internal reflection, $|\sin(\chi_{\mbox{c}})| = 1/n \approx 0.3$, has been launched. In our case, we select $|\sin{\chi}| \geq 0.5$ in order to eliminate the very short-lived rays. 
It can be seen that the scattering from CCW to CW is weaker than from CW to CCW, as in the wave calculations in Fig.~\ref{fig:dynamicsCCW}. However, the ratio of the scattering rates, corresponding to $1/|\eta|$, is below $2$, i.e., the asymmetry of scattering seems to be weaker for the ray dynamics. Note, however, that the asymmetry here depends on the initial conditions of the rays. Restricting the initial rays to regions near confined periodic ray trajectories can enhance $1/|\eta|$ to around $10$. 

\begin{figure}[ht]
\includegraphics[width=\figurewidth]{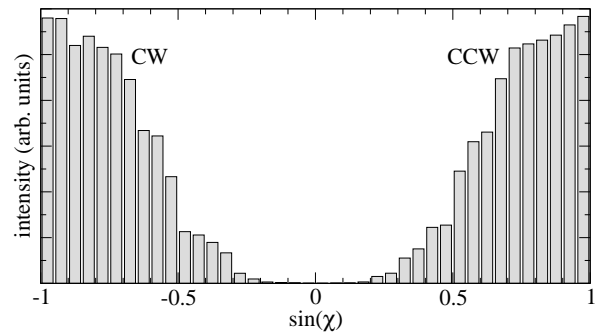}
\caption{Intensity of long-lived rays vs angle of incidence $\chi$ in the asymmetric {\limacon}. Positive (negative) $\chi$ correspond to CCW (CW) propagation direction.} \label{fig:raydistribution}
\end{figure}

Having observed this correspondence of rays and waves in terms of scattering, we show now that there is no such correspondence of rays and optical modes in terms of spatial chirality. To see this, consult Fig.~\ref{fig:raydistribution}, which is the ray analog of Fig.~\ref{fig:AMD2}(a). In total $40\,000$ rays with initially uniform distribution along the boundary of the cavity and uniformly distributed $|\sin{\chi}| \in (0.5,1)$  have been started. The quantity $\sin{\chi}$ is related to the angular momentum of a ray. In the case of the circle the relation is given by $nkR\sin{\chi} = m$ with angular momentum index~$m$~\cite{ND97}.  Positive values of $\sin{\chi}$ correspond to CCW propagation direction and negative to CW propagation direction. After time $t=50$, measured in length of ray trajectory in units of $R$, the remaining intensity of rays approaches a survival probability distribution~\cite{RLKP06}. This distribution reflects the long-time behavior of the light rays which can be compared to the properties of the long-lived modes. The survival probability distribution as function of $\sin{\chi}$ is plotted in Fig.~\ref{fig:raydistribution}. We do not observe a chirality in this distribution, i.e., the amount of CW propagating rays is with $49.8$ per cent roughly equal the amount of CCW propagating rays, which is in strong contrast to the properties of the optical modes in Fig.~\ref{fig:AMD2}(a).

Neither does the spatial chirality show up in an extended ray dynamics including first-order wave corrections such as the Goos-H\"anchen shift (GHS) and the Fresnel filtering (FF). The GHS is a lateral shift of totally reflected beams along the optical interface~\cite{GH47}, i.e.,~the points of incidence and reflection do not coincide. 
In the case of the FF~\cite{TS02,HS06,ADH08}, partial waves with angles of incidence below the critical angle for total internal reflection are (partially) refracted out of the cavity, leading to a shift $\Delta \chi$ of the partial waves between the incident and outgoing angles -- i.e., a violation of Snell's law. In the short-wavelength limit $\lambda\to 0$ the GHS and FF disappear leading to the standard ray dynamics of geometric optics. Such wave corrections have been used to explain properties of optical modes in deformed subwavelength-scale microdisk cavities~\cite{UWH08,UW10,SGS10}. 

The presented ray simulations for the asymmetric {\limacon} show that the weak asymmetry in the scattering between CCW to CW propagating rays does not lead to a chirality in the survival probability distribution. Hence, the ray dynamics cannot explain the chirality observed in the optical modes. To support this finding, in the next section we discuss another cavity geometry, for which it is rigorously proven that there is no scattering between CCW and CW propagating rays and therefore no asymmetry in the scattering of rays. Nevertheless, the modes show significant chirality and nonorthogonality. 

\section{Gutkin's billiard of constant width}
\label{sec:constantwidth}
\begin{figure}[ht]
\includegraphics[width=0.8\figurewidth]{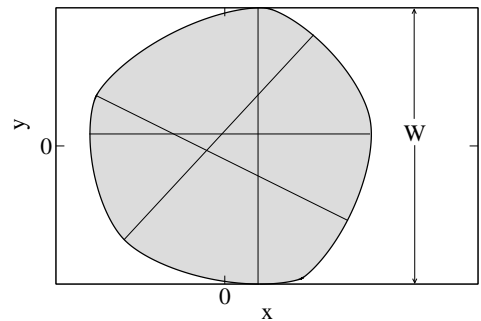}
\caption{Billiard boundary of constant width as defined in Eq.~(\ref{eq:Gutkin}) with $z(0) = (1/4-i)R$, $a_0 = R$, $a_3 = iR/8$, $a_5 = (1+i)R/4$, and $a_{2k+1} = 0$ for $k>2$. The straight lines show four possible ways (out of infinitely many) to measure the width $W$. The value of $W$ is always $2R$.}
\label{fig:Gutkinsketch}
\end{figure}

Gutkin studied a class of convex billiards (closed cavities) of constant width~\cite{Gutkin07}, i.e., for any point at the boundary the maximal distance to other points of the boundary is a constant; see, e.g., Fig.~\ref{fig:Gutkinsketch}. The (conventional and extended) ray dynamics in such billiards is characterized by a phase space which is strictly separated into two parts corresponding to CW and CCW motion, i.e., there is no scattering from CW to CCW propagating rays. Optical microcavities of constant width have been studied in the context of directional light emission~\cite{BHP04}. The parametrization of the class of boundary shapes in the $(x,y)$-plane is most conveniently given in the complex variable $z=x+iy$ 
\begin{equation}\label{eq:Gutkin}
z(\alpha) = z(0)-i\sum_{n\in \mathbb{Z}}\frac{a_n}{n+1}\left(e^{i\alpha(n+1)}-1\right)
\end{equation}
with $\alpha \in [0,2\pi)$, $a_{-n} = a^*_n$, $a_1=0$, and $a_{2n} = 0$ for $n > 0$.  Here we consider a realization without mirror symmetry and where the CW and CCW component in phase space are almost fully chaotic: $z(0) = (1/4-i)R$, $a_0 = R$, $a_3 = iR/8$, $a_5 = (1+i)R/4$, and $a_{2k+1} = 0$ for $k>2$. The constant width is $W = 2R$. This boundary curve is illustrated in Fig.~\ref{fig:Gutkinsketch}.

\begin{figure}[ht]
\includegraphics[width=\figurewidth]{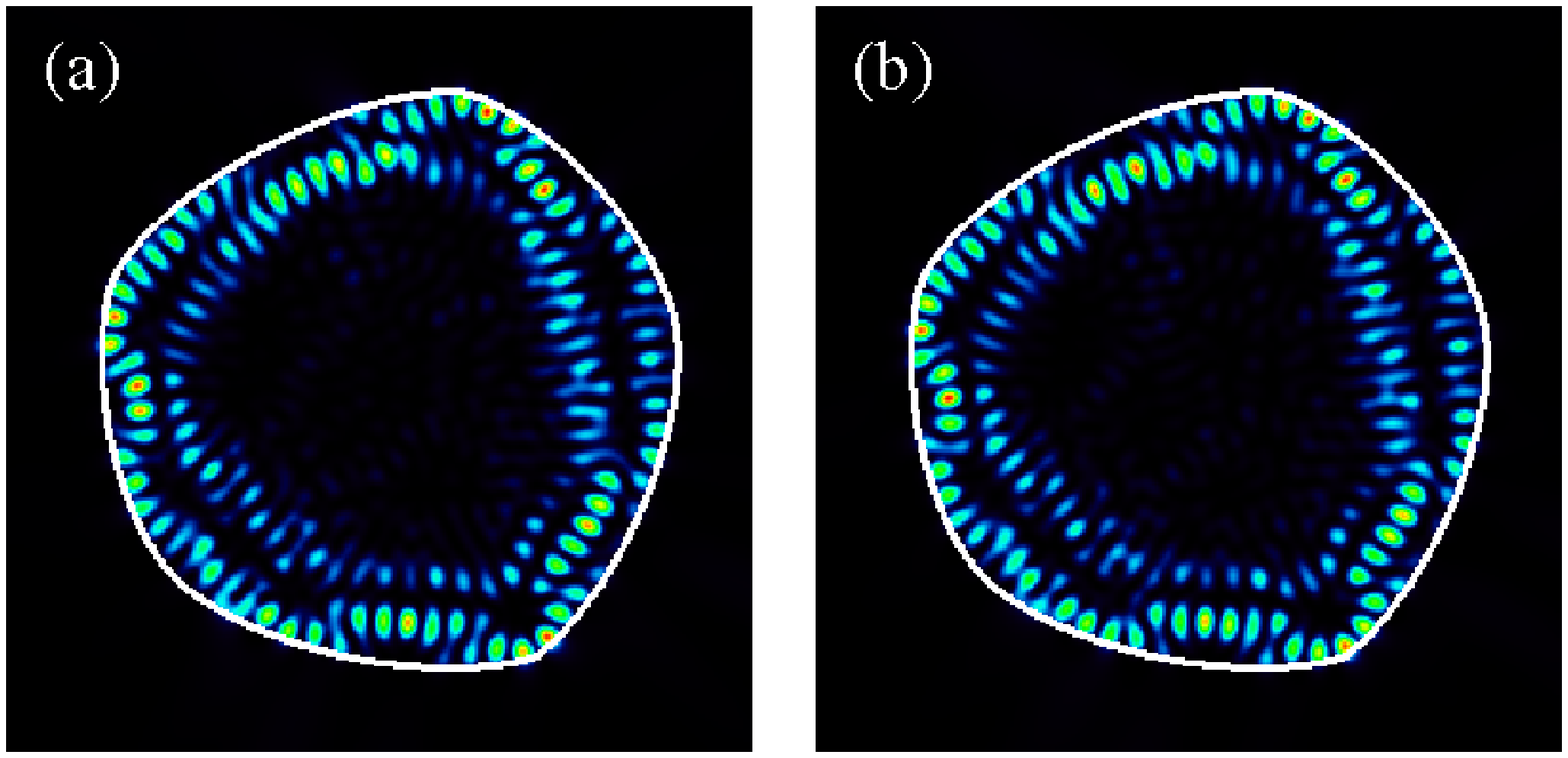}
\caption{(Color online) Intensity $|\psi|^2$ of the nearly degenerate pair of modes in Gutkin's billiard of constant width with (a) $\Omega_1 = 12.386847-i0.00561$ and (b) 
$\Omega_2 = 12.386995-i0.005695$.}
\label{fig:modeGutkin}
\end{figure}

\begin{figure}[ht]
\includegraphics[width=\figurewidth]{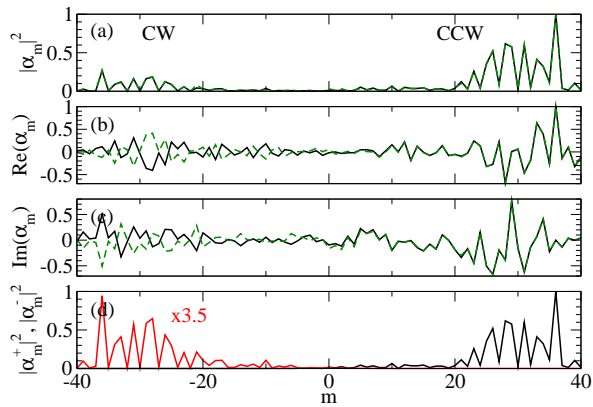}
\caption{(Color online) Angular momentum distributions $\alpha^{(1)}_m$ (black solid line) and $\alpha^{(2)}_m$ (green dashed) of the modes in Fig.~\ref{fig:modeGutkin} normalized to 1 at maximum: (a) absolute
value squared, (b) real and (c) imaginary parts, and (d) superpositions $\alpha^+_m =
 (\alpha^{(1)}_m+ \alpha^{(2)}_m)/2$ (black solid) and $\alpha^-_m
= (\alpha^{(1)}_m-\alpha^{(2)}_m)/2$ (red dashed, multiplied by a factor of 3.5).}
\label{fig:AMDGutkin}
\end{figure}

We find that the optical modes in a microcavity with such a boundary shape also appear in nonorthogonal and chiral pairs of modes. Figures~\ref{fig:modeGutkin} and~\ref{fig:AMDGutkin} show the mode pattern and its angular momentum decomposition (origin is the center of mass) of a typical pair of modes for TM polarization and refractive index $n=3.3$. The chirality turns out to be $\alpha \approx 0.726$ and $\approx 0.717$. The spatial overlap of both modes is around $0.56$. These values are in good agreement with the result from the effective Hamiltonian in Eq.~(\ref{eq:chirality2by2}). 

A proper rotation of Fig.~\ref{fig:Gutkinsketch} shows that the boundary curve is less asymmetric than the asymmetric {\limacon}. We therefore consistently observe smaller chirality $\alpha$ and overlap $S$ for the cavity geometry defined by Eq.~(\ref{eq:Gutkin}). Moreover, we find that the relative number of CW copropagating pairs is larger if compared to the asymmetric {\limacon}.

In a billiard of constant width there is no scattering from CW and CCW propagating rays and vice versa. Using this fact, we can rule out the asymmetric scattering of rays as the origin of the spatial chirality in this system. 

\section{Summary}
\label{sec:summary}
The nonorthogonality and spatial chirality of mode pairs in two asymmetrically deformed microdisk cavities, the asymmetric {\limacon} and Gutkin's cavity of constant width, has been studied. Our results indicate that the appearance of such nonorthogonal chiral pairs is a common feature of deformed microdisks which lack mirror symmetries. 
Using an effective non-Hermitian Hamiltonian we have linked these two interesting effects and explained them by the asymmetric scattering between clockwise and counterclockwise propagating waves, expressed by an asymmetry parameter~$|\eta|$. We have shown that the observation of these effects in dynamical experiments with waves depends not only on the ratio between the period related to the mode splitting~$T$ and the decay time~$\tau$ but also strongly on the asymmetry parameter~$|\eta|$. 
Finally, we have demonstrated for the considered cavities that there is no significant chirality in the survival probability distribution of rays. This is in strong contrast to the case of the spiral cavity. This observation shows that the nonorthogonality and spatial chirality is in general a wave dynamical effect.

It remains an interesting question for future research to ask for the size of the chiral effects in the case of circular-shaped disks with surface roughness arising from the inevitable imperfections in the fabrication process. In such a case we also expect the appearance of nonorthogonal pairs of copropagating modes. But the nonorthogonality and the chirality might be small for the case of weak surface roughness that can be achieved nowadays in state-of-the-art experiments. Moreover, averaged over many mode pairs the chirality should be close to zero due to the random character of the boundary profile.

We believe that our results are not only important for deformed microdisks but also for other types of optical microcavities (microspheres and microtoroids) and for open quantum (wave) systems in general.

\acknowledgments 
We thank Julia Unterhinninghofen and Sang Wook Kim for discussions. Financial support from the DFG research group 760 and DFG Emmy Noether Programme is acknowledged.  


\end{document}